\documentclass{emulateapj}

\shorttitle{NGC1277: a massive compact relic galaxy in the nearby Universe}
\shortauthors{I. Trujillo et al.} 

\def\gsim{ \lower .75ex \hbox{$\sim$} \llap{\raise .27ex \hbox{$>$}} }
\def\lsim{ \lower .75ex \hbox{$\sim$} \llap{\raise .27ex \hbox{$<$}} }

\begin{document}
\title{NGC1277: a massive compact relic galaxy in the nearby Universe}

\author{Ignacio Trujillo$^{1,2}$, Anna Ferr\'e-Mateu$^{3}$, Marc Balcells$^{4}$, Alexandre Vazdekis$^{1,2}$ and
 Patricia S\'anchez-Bl\'azquez$^{5}$} 
\affil{$^{1}$Instituto de Astrof\'{\i}sica de Canarias,c/ V\'{\i}a L\'actea s/n, E38205 - La Laguna, Tenerife, Spain}
\affil{$^{2}$Departamento de Astrof\'isica, Universidad de La Laguna, E-38205 La Laguna, Tenerife, Spain}
\affil{$^{3}$Subaru Telescope, 650 North A'ohoku Place, Hilo, 96720, Hawaii, USA}
\affil{$^{4}$Isaac Newton Group of Telescopes, E-38700 Santa Cruz de La Palma, Canary Islands, Spain}
\affil{$^{5}$Departamento de F\'isica Te\'orica, Universidad Aut\'onoma de Madrid, E28049, Cantoblanco, Madrid, Spain }
\email{email:trujillo@iac.es}

\begin{abstract}

As early as 10 Gyr ago, galaxies with more than 10$^{11}$ M$_{\odot}$ in stars already existed. While most of
these  massive galaxies must have subsequently transformed through on-going star formation and mergers with other
galaxies, a small fraction ($\lesssim$0.1\%) may have survived untouched till today. Searches for such relic
galaxies, useful windows to explore the early Universe, have been inconclusive to date: galaxies with masses and
sizes like those observed at high redshift (M$_\star$$\gtrsim$10$^{11}$ M$_{\odot}$; R$_e$$\lesssim$1.5 kpc) have
been found in the local Universe, but their stars are far too young for the galaxy to be a relic galaxy. This
paper explores the first case of a nearby galaxy, NGC1277 (at a distance of 73 Mpc in the Perseus galaxy
cluster), which fulfills many criteria to be considered a relic galaxy. Using deep optical spectroscopy, we
derive the star formation history along the structure of the galaxy: the stellar populations are uniformly old
($>$10 Gyr) with no evidence for more recent star formation episodes. The metallicity of their stars is
super-solar ([Fe/H]=0.20$\pm$0.04 with a smooth decline towards the outer regions) and alpha enriched
([$\alpha$/Fe]=0.4$\pm$0.1). This suggests a very short formation time scale for the bulk of stars of this
galaxy. This object also rotates very fast (V$_{rot}$$\sim$300 km/s) and has a large central velocity dispersion
($\sigma$$>$300 km/s).  NGC1277  allows the explorations in full detail of properties such as the structure,
internal dynamics, metallicity and initial mass function at $\sim$10-12 Gyr back in time when the first massive
galaxies were built.

\end{abstract}


\section{Introduction}

We define a relic galaxy as an object that has been formed in the early phases of the Universe (i.e. z$>$2) and
which has remained unaltered (i.e. without significant gas or stellar accretion) since its initial formation.
Finding an object with these characteristics in the local Universe would allow a detailed study of its assembly
history, and significantly advance our understanding of the first steps in galaxy formation. The question then is
how to identify one of these objects in our neighborhood. One obvious characteristic that these relic galaxies
should have is that they must to be old along their entire structure and with no signs of star formation after
its formation. In other words, their star formation history (SFH) should peak at the beginning of the Universe
and not show any activity later on. 

Among  present-day galaxies, the most promising candidates to be relics are the massive ellipticals. 
Their SFHs are compatible with little or negligible star formation \citep[e.g.][]{1997ApJS..111..203V}  after their
initial formation. However, their structures are very different \citep[i.e. their
sizes have grown by a factor of $\sim$4;][]{2007MNRAS.382..109T} compared to the size of their likely
progenitors at z$>$2, implying that they have assembled a significant fraction of their stars in the last 10
Gyr. For this reason, if we want to find present-day galaxies which have remained unaltered since their initial
formation at z$>$2, we have to search for current
old massive galaxies which have the same structural properties than the population of massive galaxies at
high-z. This is, finding  old ($>$10 Gyr), massive (M$_\star$$\gtrsim$10$^{11}$ M$_{\odot}$) and compact
(R$_e$$\lesssim$1.5 kpc) galaxies in our vicinity. 

Recently, \cite{2013ApJ...773L...8Q} estimated the number density and the expected fraction of massive relic
galaxies in our nearby (z$\lesssim$0.1) Universe. They found that only $\sim$0.1\% of current massive
galaxies are expected to have accreted less than $\sim$10\% of their stars after their formation. This
implies that the today number density of massive relics should be $\sim$10$^{-6}$Mpc$^{-3}$. These figures
are small, but thanks to large area surveys like the Sloan Digital Sky Survey (SDSS) we should be able to
find $\sim$60 massive relics down to z=0.1 (in the 8032 square degree covered by the spectroscopic SDSS
Legacy DR7). Searches for massive compact galaxies in the nearby Universe have found some (although less than
expected) candidates to be relic objects
\citep[e.g.][]{2009ApJ...692L.118T,2010ApJ...712..226V,2010ApJ...720..723T,2013ApJ...762...77P} but it turned
out that these galaxies are surprisingly young \citep[$\sim$2 Gyr;][]{2012MNRAS.423..632F}. It is unclear
whether the absence of relics in the SDSS Legacy  footprint could be related to spectroscopic incompleteness
in some particular areas of the sky. For instance, in rich galaxy clusters, the spatial proximity of the
galaxies could have prevented a proper spectroscopic coverage of the targets. 

The object that we explore in this paper, NGC1277, is located at 73 Mpc in the Perseus galaxy cluster. This
cluster is outside the SDSS DR7 Legacy coverage. However, this field was targeted by an exploratory survey
for preparing the SDSS SEGUE program. NGC1277 has recently received a lot of attention  due to the claim by
\cite{2012Natur.491..729V} that this object has an extraordinary large supermassive black hole (SMBH). 
Interestingly, NGC1277 is massive (M$_\star$=1.2$\pm$0.4$\times$10$^{11}$ M$_{\odot}$) and compact (R$_e$=1.2
kpc), two of the three properties we request for a massive galaxy to be a relic. This letter explores the
third condition for being a relic galaxy and presents a detailed analysis of the SFH of NGC1277 along its
global structure. The SDSS spectrum of this galaxy is insuficient for this purpose as it only covers its
inner 0.5 kpc. We use our own very deep spectroscopy to perform a stellar population analysis out to
$\sim$3R$_\mathrm{e}$. NGC1277 is uniformly old, which makes it a strong candidate to be a relic galaxy.

Throughout we adopt a standard cosmological model with the following parameters: $H_0$=70 km s$^{-1}$ Mpc$^{-1}$,
$\Omega_m$=0.3 and $\Omega_\Lambda$=0.7. The redshift adopted here for NGC1277, z=0.0169, corresponds to a
galaxy distance of 73.3 Mpc and to a spatial scale of 344 pc/arcsec. 

\section{Data}

\subsection{HST imaging}

The stellar mass density profile of NGC1277 as well as the high resolution imaging of the galaxy  have been
obtained using the WFC ACS  F625W (Sloan r) filter from the NASA/ESA Hubble Space Telescope (HST) archive.
These observations are associated with the program GO:10546 (PI: Fabian). The pixel scale of this image is
0.05\arcsec\ and the Point Spread Function Full Width at Half Maximum is 0.1\arcsec\ . The
photometric zeropoint is 25.893 (AB system). Figure 1 shows an image of this galaxy in the F625W band.

\subsection{Deep long-slit spectroscopy}

We obtained deep long-slit spectra employing the ISIS spectrograph on the 4.2m William Herschel Telescope. The
blue grating R300B was centered at 5300$\rm\AA{}$ with a $\times$2 binning in the spatial direction and the
1\arcsec\ slit was placed along the galaxy's major axis (P.A.\,=\,96 degrees). This setup provided a spectral
resolution of 3.4\,$\rm\AA{}$. Seeing was  0.7\arcsec\ during the observations.  Six exposures of 30 minutes
were taken, giving a total time on source of 3h. Two spectrophotometric stars were observed with the same
configuration to correct the shape of the spectra.

Data reduction was performed with REDUCEME \citep{1999PhDT........12C}, a package optimized for
long-slit spectroscopy that allows a parallel treatment of the scientific data and the errors propagated
through the process. This reduction process included a bias subtraction, flat-fielding, cosmic-ray removal,
C- and S-distortion correction, wavelength calibration, sky subtraction and flux calibration. The result is
a high quality spectrum (averaged central S/N$>$150) that covers the wavelength range
$\lambda\lambda$\,3400-6200\,$\rm\AA{}$. This coverage is optimal as it encompasses most of the
line-strength indices employed in  stellar population analysis and is also wide enough for the
full-spectral-fitting technique in which this exercise is mostly based on. 

\section{Analysis}

One of the conditions that we demand of a relic galaxy is to show no signatures (i.e. tidal tails,
asymmetries, etc) of present or past interactions. In the case of NGC1277 this analysis is complicated due to
the presence of two  galaxies aligned near our object (Fig.1, left panel). Consequently, in order to study
potential faint evidence of interactions, we have substracted the light contamination from those galaxies. We
have modeled and subtracted the light of the two intervening galaxies using the \textsc{IRAF} task
\textsc{ELLIPSE} \citep[][]{1987MNRAS.226..747J}. 

The clean image of NGC1277 is shown in Fig. 1, right panel. We see no tidal signatures down to the surface
brightness limit of this image ($\sim$26.8 mag/arcsec$^2$; r-band). In fact, NGC1277 looks very regular along
its entire structure. A conservative estimation suggests that the above surface brightness limit implies that
any potentially "hidden" disrupted satellite around NGC1277 would contribute with less than 2\% to the mass
of the galaxy.  For this reason, we conclude that NGC1277 does not present signs of being currently (or in
its inmediate past) accreting external stars.

\subsection{Stellar mass density profile}

The second evidence  to conclude that NGC1277 has the structural characteristics of the primitive massive
galaxies is given by its stellar mass density profile. Although our galaxy is clearly elongated, the  mass
density profiles of the galaxies which we have used in our control samples, both at low and high-z (see next
paragraph), were obtained using circular apertures. For this reason, to make a fair comparison with those
control galaxies, we have obtained the  surface brightness profile of our galaxy using circular apertures. 
Once we have the F625W (r-band) surface brightness profile, we have created the stellar mass density profile
of the galaxy assuming that there are not stellar population gradients through its entire structure. This
approximation is  reasonable according to the information provided by the deep spectra we will show in the
next subsection. Using the age and metallicity from our analysis of the galaxy spectra (12 Gyr and 0.2 dex
respectively; see below), we have obtained a M/L of 4.438 (r-band) which we use to build the stellar mass
density profile (see Fig. 2).

In the left panel of Fig. 2, we show the comparison of the stellar mass density profile of NGC1277 with the
stellar mass density profiles of "normal-sized" massive galaxies from the SDSS survey. To build the stellar
mass density profiles of the "normal-sized" galaxies, we took the structural parameters (S\'ersic index n,
effective radius R$_e$, and stellar mass M$_\star$) of all the galaxies in the NYU catalog
\citep{2005AJ....129.2562B} with 0.8$<$M$_\star$$<$1.2$\times$10$^{11}$M$_\odot$  and 0.08$<$z$<$0.12.  To
facilitate the comparison with our profiles, we divided the NYU galaxies into two different categories:
disk-like (n$<$2.5) and spheroid-like (n$>$2.5). We find that the average disk-like massive galaxy within
the NYU sample at those redshifts has M$_\star$=0.92$\times$10$^{11}$M$_\odot$, n=2.2, and R$_e$=5.0 kpc.
On the other hand, the average spheroid-like object has M$_\star$=0.95$\times$10$^{11}$M$_\odot$, n=4.0,
and R$_e$ = 4.7 kpc. Once we obtained these average galaxy profiles, the representative regions of each
galaxy category were build using all the galaxies in the NYU sample within the above stellar mass range and
redshift interval whose central stellar mass densities were within 68\% of the distribution centered around
the mean value.

Neither present-day massive "normal-sized" spiral-like nor elliptical-like  galaxies 
have similar stellar mass density profiles to the one shows by NGC1277.  NGC1277 has a denser profile in
the inner 1 kpc (being a factor 2-3 denser than the present-day densest elliptical galaxies of the same
stellar mass). In the outer region, beyond 4 kpc, NGC1277 is underdense compared to the normal population
of massive galaxies. However, when we compare (see Fig. 2 right panel) the stellar mass density profile of
NGC1277 with the massive compact galaxies found at high-z  \citep{2012ApJ...749..121S} or with the young
compact massive galaxies  found at z$\sim$0.15 \citep{2012ApJ...751...45T}, the agreement between both
profiles is remarkable. We conclude that the detailed structural properties of NGC1277 are equivalent to
the ones found in the primitive Universe for massive galaxies with similar stellar mass.

\subsection{Stellar population properties and SFH}

The final evidence for considering that NGC1277 is a relic galaxy comes from a detailed analysis of its
spectra along its major axis (see Fig. 3). In the upper panel of this figure, we present the spectra of the
galaxy at different radial distances. The depth of our exposure allows us to explore the stellar population
properties of our galaxy down to 3R$_e$ with a S/N above 20 \citep[our limit to ensure that we have enough quality to
get reliable SFHs;][]{2013arXiv1307.0562C}.

The SFHs of NGC1277 along its radial distance have been obtained  using a full-spectral-fitting approach.  We
have employed the extended version of MILES stellar population synthesis models \citep{2010MNRAS.404.1639V} 
MIUSCAT \citep{2012MNRAS.424..157V,2012MNRAS.424..172R} and we have fed the full-spectral-fitting code  {\tt
STARLIGHT} \citep{2005MNRAS.358..363C} with them in order to recover the SFH of this galaxy.  The models
cover a wide range of both ages (0.1-17.8\,Gyr) and metallicities (-1.71 to 0.22) and they also allow for
variations on the IMF slope and shape. For the purpose of this exercise, we have focused on the standard
assumption of a Kroupa Universal IMF, although we have also studied the impact on changing the IMF
\citep{2013MNRAS.431..440F} according to the velocity dispersion, as recently claimed by several studies
\citep[e.g.][]{2013MNRAS.429L..15F,2013MNRAS.433.3017L} and we have found that our results remain unaltered.
In addition, the same study was carried out with another different spectral-fitting code, {\tt STECKMAP}
\citep{2006MNRAS.365...46O} to ensure the robustness of the method, rendering similar results.

We therefore derive the mean age and mean metallicity and the  SFH for each one of the radial binnings out to
3R$_e$  (both luminosity and mass-weighted). Figure\,3 lower panels show the derived mass-weighted SFH for 4
of the apertures. It is clear that all the stars in this object were formed at high-redshift, being most of
them older than 10\,Gyr. In fact, this remains true even out to large galactocentric distances, as we present
in Figure\,4, where the radial profiles for the mass-weighted age (top) and metallicity (middle panel) are
shown. According to the analysis done with {\tt STARLIGHT}, NGC\,1277 is an old object ($\sim$12\,Gyr) with a
high total metallicity down to at least 3R$_e$.  This corresponds to a radial distance which encloses 85\% of
the total light of this object \citep{2001MNRAS.326..869T} assuming a global S\'ersic index of n=2.2 for this
galaxy \citep{2012Natur.491..729V}.  In addition to the SFH, age and metallicity profiles obtained with a
full-spectral-fitting approach, the middle and lower panel in Fig. 4 presents the radial variation of the
metallicity and the $\alpha$/Fe parameter, which is a good estimator of how fast a galaxy forms its stars
\citep{1992ApJ...398...69W} using line strength indices. This metallicity profile has been estimated using a
hybrid approach which combines the luminosity weighted age derived from {\tt STARLIGHT} versus the [MgFe]'
index. This metallicity gradient shows similar values in the inner region than the one purely based on
STARLIGHT. Both metallicity profiles seems to indicate a smooth decline towards lower metallicities in the
outer parts NGC1277.

To estimate the $\alpha$/Fe gradient, since the MIUSCAT models are scaled solar around the solar metallicity,
we employ the approach shown in \cite{2013MNRAS.433.3017L}. We have first derived two independent metallicity
estimates from the pair of spectral indices H$\beta$-Mgb$_{5177}$ and H$\beta$-$<$Fe$>$* \citep[a combined Fe
index;][]{1993PhDT.......172G}, Z$_{Mg}$ and Z$_{Fe}$ respectively. The difference between these two
metallicities is a good solar proxy [Z$_{Mg}$/Z$_{Fe}$] and has been shown to tigthly correlate with the
$\alpha$-enhanced SSP models of \cite{2011MNRAS.412.2183T}. We have found [Z$_{Mg}$/Z$_{Fe}$]$\sim$0.7 in our
spectra, which translates into [$\alpha$/Fe]$\sim$0.4. Note that an extrapolation has been made when
employing the model grid corresponding to the Mgb index \citep[see also Fig. 5 in][]{2013MNRAS.433.3017L},
and therefore these values should be taken with some caveats. Nonetheless, such very high values indicate
that the bulk of the stellar populations of this object, which are very old, was formed in a very short
period of time, almost resembling a single-burst event. In fact, according to the calibration provided by
\citet[][their Eq. 2]{2011MNRAS.418L..74D}  between  the time needed to form half of the final stellar mass
of the galaxy, T$_{M/2}$, and $\alpha$/Fe, NGC1277 formed its mass in less than a few hundred Myrs. For
comparison, a normal-sized massive galaxy with $\alpha$/Fe=0.25, has T$_{M/2}$=1.4 Gyr. In other words,
NGC1277 seems to have formed its stellar mass much faster than "normal-sized" galaxies with equal
mass. Such short time scales suggest star formation rates as high as $\sim$1000 M$_{\odot}$/yr for forming
the bulk of the stars of NGC1277. These high star formation rates have been measured in massive high-z
galaxies \citep{2013Natur.496..329R}.

\begin{figure*}
\includegraphics[scale=1]{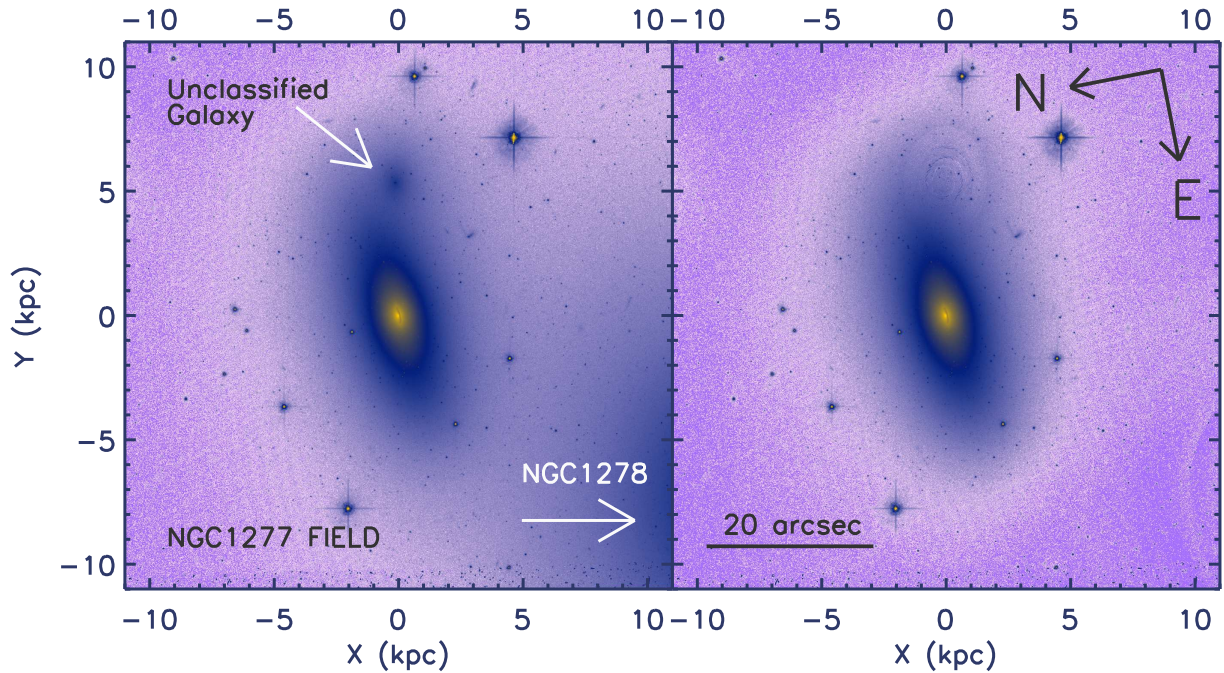}

\caption{The neighborhood of NGC1277 as seen by the HST F625W filter. The left panel shows the two closest
galaxies whose light contaminate  NGC1277.  The right panel shows NGC1277 after the
subtraction of the contaminant light.  The results indicates that NGC1277 is rather symmetric with no
distortions neither bright tidal streams surrounding it.}

\label{fig1}
\end{figure*}

 \section{Results and Discussion}

We have presented three strong evidences that suggest that NGC1277 is a relic galaxy. It is massive and
compact, with no signature of interactions, it has exactly the same stellar mass density profile than the
massive compact galaxies at high-z, and finally its stellar populations are compatible with having been
formed very fast and more than 10 Gyr ago. Consequently, it is worth exploring its other properties to
learn about the physical conditions at the formation of the massive galaxies in the early Universe.

No doubt, one of the most remarkable claims about NGC1277 is the presence of an over-massive blackhole in its
center \citep{2012Natur.491..729V}. The exact mass of this SMBH is controversial, with claims that range from
$\sim$2$\times$10$^{10}$M$_\odot$ to more modest  (but still a factor of $\sim$4  larger than the expectation for
the mass of this galaxy) values as $\sim$2$\times$10$^{9}$M$_\odot$ \citep{2013MNRAS.433.1862E}. One is tempted
to claim that the reason why this SMBH is much larger than what it is expected  is the absence of further growth
in mass of its host galaxy after its initial assembly. In fact, an object with the characteristics of NGC1277 is
expected to double its mass since z$\sim$2 due to major and minor (dry) merging  \citep{2010ApJ...709.1018V}.
This accreted stellar mass is mainly deposited in the outer region of the galaxies without feeding with new gas
the central SMBH. For this reason, if NGC1277 had followed the normal growth path expected for this type of
galaxies, it would have had a more "normal" SMBH. If this picture is correct \citep[see
also][]{2013ARA&A..51..511K}, it seems reasonable to suggest that the SMBHs (at least for the most massive
galaxies) were formed together with the bulk of the stars of their host galaxies in a very fast collapse at
high-z. After that, the SMBHs have remained unchanged in mass while the mass of the host galaxies have continued
growing by successive merging. It would be worth exploring whether the peculiar evolutionary path
followed by NGC1277 was dictated by its location in a rich cluster as Perseus.

The morphology and internal dynamics of NGC1277 are also interesting. Visually, NGC1277 has been classified
as a peculiar S0 \citep{1994AJ....108.2128C}. In fact, its elongated shape resembles such morphology. It is
worth noting that  high ellipticity is common among massive compact galaxies, both at high-z
\citep[e.g.][]{2011ApJ...730...38V,2013MNRAS.428.1460B} and at
z$\sim$0.15 \citep{2012ApJ...751...45T}. Finally, in relation to the dynamics of NGC1277 it is worth
mentioning the high central velocity dispersion (333 km/s) as well as its fast rotation ($\sim$300 km/s)
measured along its major axis \citep{2012Natur.491..729V}.  To go further in the dynamical analysis, and also
to address better the  morphology of NGC1277, is necessary to explore the dynamics of this object with 3D
spectroscopy. At this moment, with the information along the major axis, we can only speculate. If this
galaxy was in fact formed in a very fast event, we can think that the dynamics of its most inner region could
resemble the turbulent and chaotic motions of the shocks of enormous cold flows triggering the star formation
in its center. Its fast rotation also could be related to the compact structure of NGC1277. The angular
momentum conservation could have transformed an  original modest rotational velocity of the infalling gas
cloud into the high values we observe now for NGC1277 outer stars. 

Finally, if the theoretical predictions by \cite{2013ApJ...773L...8Q} are correct, one would expect to find
only a single relic galaxy every 10$^6$ Mpc$^3$. It turns out that this number is very close to the volume
enclosed by a sphere with a radius  of 73 Mpc ($\sim$1.6$\times$10$^6$ Mpc$^3$). This again points out to the
unique nature of NGC1277.

\begin{figure*}
\includegraphics[scale=1]{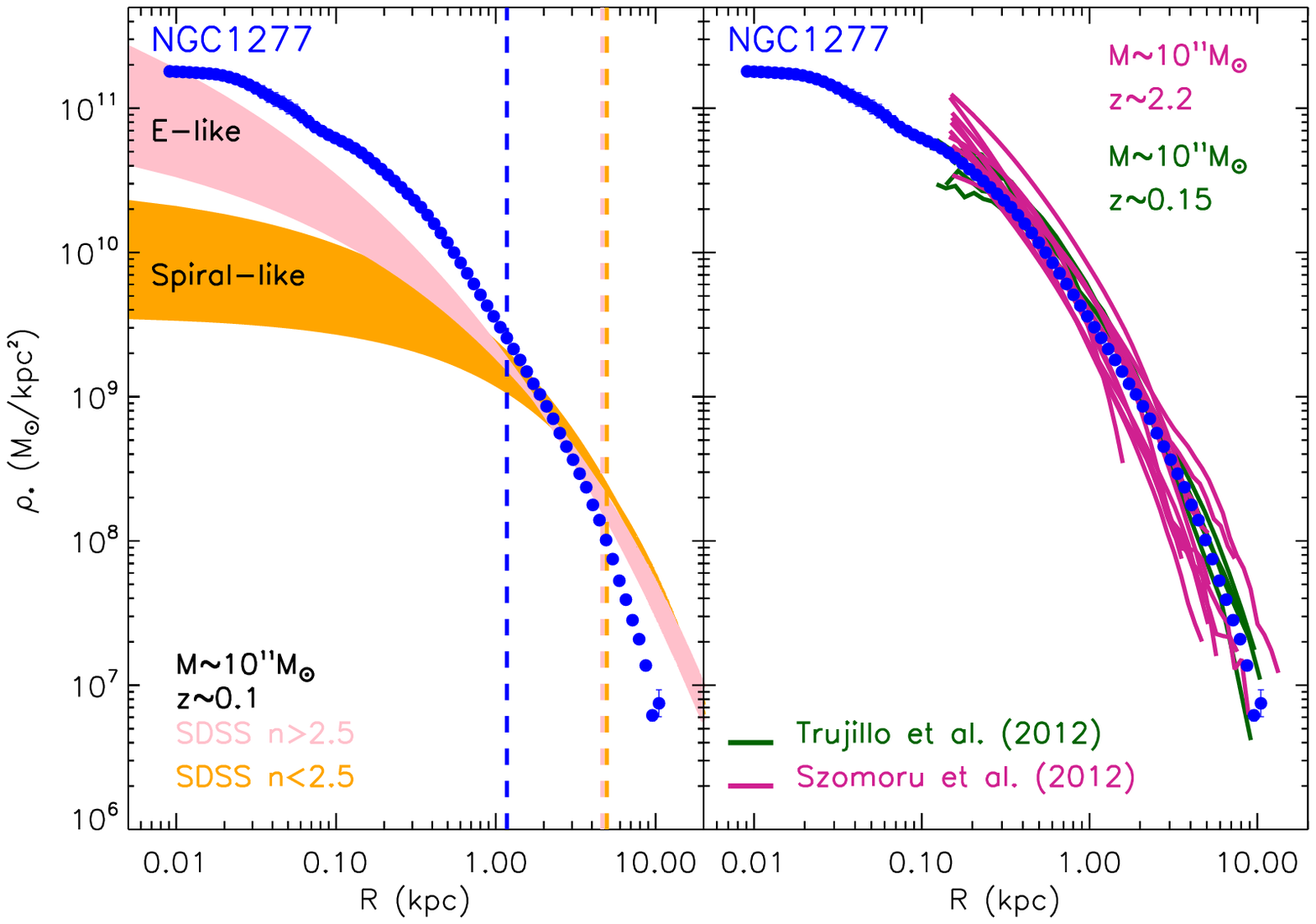}

\caption{ The circularized stellar mass density profile of  NGC1277.  The left panel  shows the comparison
between the stellar mass density profile of NGC1277 and similar mass but  normal-sized SDSS galaxies.  The
vertical dashed lines indicate the position of the effective radii for NGC1277 (blue line; 1.2 kpc) and
SDSS massive ellipticals (pink line; 4.7 kpc) and spirals (orange line; 5 kpc). The right panel shows the
comparison of the stellar mass density profile of NGC1277 with similar mass compact galaxies at high-z and 
young compact massive objects at z$\sim$0.15.} 

\label{fig2}
\end{figure*}

\begin{figure*}
\includegraphics[scale=1]{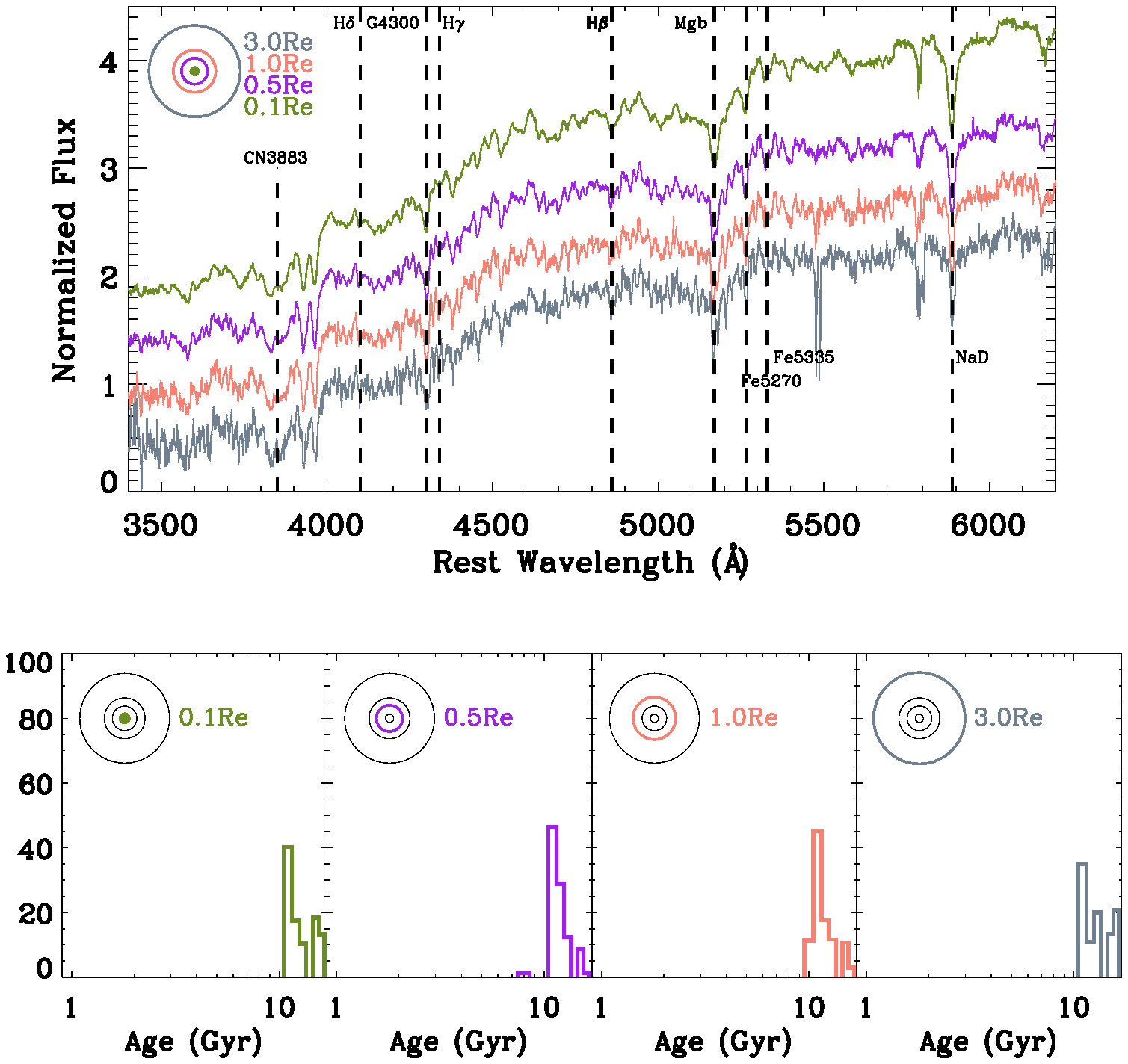}

\caption{Spectral energy distribution of  NGC1277 at different radial distances. The spectra are shown in
arbitrary units, with the flux normalized per unit wavelength and shifted for clarity. The position of
several relevant absorption lines are indicated with vertical lines. The bottom panels show the SFHs
derived with {\tt STARLIGHT} at different radial distances. This represents the fraction of mass created at
each epoch.} 

\label{fig3}
\end{figure*}

\begin{figure}
\includegraphics[scale=.7]{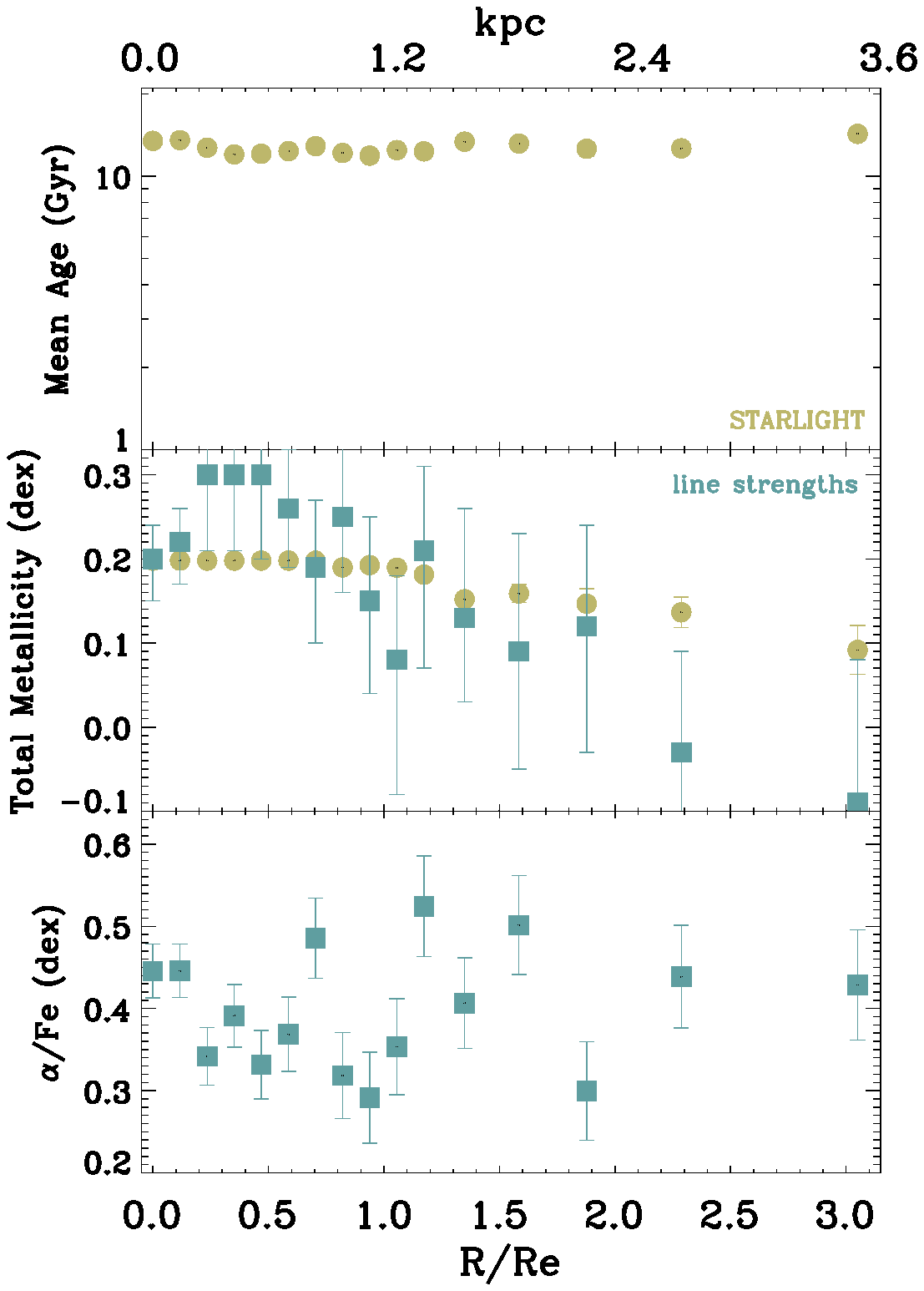}

\caption{Age, metallicity and $\alpha$/Fe profiles of NGC1277. The figure shows the mean mass-weighted age
and total metallicity derived  using {\tt STARLIGHT}. An independent measured of the metallicity and
$\alpha$/Fe was obtained from an indices analysis. The profiles show a small change of the stellar population
properties of this relic galaxy across its structure out to 3R$_e$. It seems that the entire galaxy was
formed in a unique, very fast event, which produced the high $\alpha$/Fe abundances.} 

\label{fig4}
\end{figure}

\acknowledgments We thank the referee for his/her constructive comments. This work benefited from interesting
discussions with Jes\'us Falc\'on-Barroso and Francesco La Barbera. This article is based on observations made with the
WHT operated on the island of La Palma by the Isaac Newton Group of Telescopes in the Spanish Observatorio
del Roque de los Muchachos. This research has been supported by the Spanish Ministerio de Econom\'ia y
Competitividad (MINECO; grants AYA2010-21322-C03-02 and AYA2009-11137). AFM acknowledges the Japan Society
for the Promotion of Science (JSPS) Grant-in-Aid for Scientific Research (KAKENHI) Number 23224005.

\end{document}